\documentclass[12pt,twoside]{article}
\usepackage{cmp2e,epsfig}

\title[Criticality of the random-site Ising model]{
  Criticality of the random-site Ising model: Metropolis,
  Swendsen-Wang and Wolff Monte Carlo algorithms
}

\author[D. Ivaneyko, J. Ilnytskyi, B. Berche, Yu. Holovatch]{
  D. Ivaneyko\refaddr{LUniv},
  J. Ilnytskyi\refaddr{ICMP},
  B. Berche\refaddr{Nancy},
  Yu. Holovatch\refaddr{ICMP,Linz,LUniv}
 }

\addresses{
  \addr{LUniv} Ivan Franko National University of Lviv,
               79005 Lviv, Ukraine
  \addr{ICMP}  Institute for Condensed Matter Physics,
               National Acad. Sci. of Ukraine,\\
               79011 Lviv, Ukraine
  \addr{Nancy} Laboratoire de Physique des Mat\'eriaux,
               Universit\'e Henri Poincar\'e 1,\\
               54506 Vand\oe uvre les Nancy Cedex, France
  \addr{Linz}  Institut f\"ur Theoretische Physik,
               Johannes Kepler Universit\"at Linz,\\
               4040 Linz, Austria
}
\begin{document}

\maketitle

\begin{abstract}
We apply numerical simulations to study the criticality
of the 3D Ising model with random site quenched dilution.
The emphasis is given to the issues not being discussed
in detail before. In particular, we attempt a comparison
of dif\-ferent Monte Carlo techniques, discussing
regions of their applicability and advantages/disadvantages
depending on the aim of a particular  simulation set.
Moreover, besides evaluation of the critical indices
we estimate the universal ratio $\Gamma^+/\Gamma^-$
for the magnetic susceptibility critical amplitudes. Our estimate
$\Gamma^+/\Gamma^- = 1.67 \pm 0.15$ is in a good agreement
with the recent MC analysis of the random-bond Ising model
giving further support that both random-site and random-bond
dilutions lead to the same universality class.\\
\pacs 61.43.Bn, 64.60.Fr, 75.10.Hk
\end{abstract}

\section{Introduction}

Since its introduction in the early 1920-ies the Ising model
serves as a paradigm to study criticality of interacting
many-particle systems where the single particle can be in two
possible states. Such a general problem statement
supported in the 1960-ies by the universality and scaling
hypothesis led to the present situation, when the Ising
model is used to
explain criticality and scaling in such different fields as
ferromagnetism and binary mixtures from the one side or networks and
text series from the other side. Subsequently, the random-site
Ising model is of primary importance to understand the influence
of the structural disorder on criticality. And this explains
a lot of theoretical, experimental and computational efforts
invested so far to shed light on the phenomena happening there.

In particular, the three-dimensional (3D) Ising model
subject to a weak dilution by non-magnetic impurities
changes its universality class \cite{Harris}.
Already this phenomena called for its
explanation and verification. We address the reader to recent
reviews \cite{Pelissetto02,Folk03} where theoretical, experimental
and computational
data is collec\-ted and compared. In our study, we use the Monte Carlo
(MC) approach and address the comparison of three different
simulational
techniques analysing critical behaviour of the 3D quenched diluted
Ising model. Some of the outcome is obvious, but we think it is
instructive to perform such simulations now, when certain MC tools are
sometimes discarded for not obvious reasons. Moreover, a part of our
simulations concerns the critical amplitude ratios of the
3D random-site Ising model -- a question which was not
intensively analysed so far, and where a disagreement between the
MC and theoretical predictions exists.

It is our big pleasure and honour to contribute this paper to the
Festschrift devoted to Reinhard Folk's 60th birthday: his contribution
to the theory of phase transitions in general and to the problem we
consider here is hard to be overestimated.

\section{Simulation details and results}\label{II}

The Hamiltonian of the random-site Ising model has the following
form
\begin{equation}
H = -J\sum_{\langle ij\rangle }c_ic_jS_iS_j, \label{Ham}
\end{equation}
where $\langle ij\rangle $ denotes the summation over the nearest
neighbour sites of 3D simple cubic lattice, $J$ is the interaction
constant, $c_i=1$ if the $i$-th site is occupied by spin and
$c_i=0$ otherwise, the Ising spins $S_i$ take two values $\pm 1$.
Occupied sites ($c_i=1$) are considered to be randomly distributed
and quenched in a fixed configuration. For every  observable,
discussed below, first the Boltzmann average with respect to the
spin subsystem is performed for the fixed site configuration,
subsequently the averaging  over different disorder realisations
(configurational average) is performed. We will use the following
notations, the number of all sites is $N=L^3$ and the number of
sites carrying a spin is $N_p$. The concentration of spins is
defined therefore as $p = N_p/N$.

Several separate sets of simulations are performed in this study
with the aim to investigate different characteristics of the
model. In all the simulations the concentra\-tion of spins was taken
fixed and equal to $p=0.85$. For the 3D random site Ising model,
at spin concentration $p\sim 0.8$ the correction-to-scaling terms
appear to be particularly small \cite{Ballesteros98}. Therefore,
for the concentration chosen we do not take these terms into
account in our further analysis. The simulations were performed on
a set of following lattice sizes $L=10,12,16,24,32,48,64,96$ with
periodic boundary conditions. The other details, particularly the
number of disorder realisations, the simulation lengths and the
algorithms employed have been chosen depending on the aim of
particular simulation set.

The aim of the first set of simulations is to estimate the
critical temperature of the model at different $L$. Due to the
finite-size scaling theory \cite{Barber}, the finite system of
linear size $L$ will demonstrate an evidence of a critical
behaviour at a certain temperature $T_c(L)$ which differs from the
critical temperature of the infinite system $T_c(\infty)$
\cite{FerLand}
\begin{equation}\label{Tc}
T_c(L) = T_c(\infty) + aL^{-1/\nu} + \ldots,
\end{equation}
where the correction-to-scaling terms have been omitted. To find
$T_c(L)$ the following procedure was followed. The initial
configuration was prepared by scattering diluted sites randomly on
the lattice and the original orientations of spins were chosen
randomly. First, the runs on the smallest lattice size $L=10$ were
performed on $10^2$ disorder realisations during $5\cdot10^5$ MC
sweeps using Metropolis algorithm \cite{Metr}. The preliminary
estimate for the critical temperature was taken from the
mean-field approximation:
\begin{equation}
T^{MF}_c = p\cdot T^{pure}_c,
\end{equation}
where $T^{pure}_c$ is the critical temperature of the pure 3D Ising
model \cite{FerLand}. During the simulations the histograms for the
potential energy (\ref{Ham}) and for the absolute value of the
order parameter
\begin{equation}
M = \frac{1}{N_p}|\sum_{i=1}^{N}c_iS_i|
\end{equation}
were built and then, using the histogram reweighting technique
\cite{FerSw}, the temperature region around $T^{MF}_c$ was explored.
For a given disorder realization, the susceptibility was
evaluated according to the fluctuational relation
\begin{equation}
\chi = KN_p(\langle M^2\rangle - \langle M\rangle^2),
\end{equation}
where $\langle \ldots\rangle$ denotes Boltzmann averaging and
$K=\beta J=J/k_B T$ is the dimensionless coupling. The
temperatures where $\chi$ has a maximal value were then averaged
over all disorder realisations (hereafter, such configurational
averaging will be denoted by $\overline{\vphantom{L}\ldots}$) and
used as the working estimate for the critical temperature
$T_{sim}(L)$ for $L=10$. This temperature was also used as the
preliminary estimate for the critical temperature for the next
lattice size $L=12$. Again, for $L=12$ we performed short runs on
$10^2$ disorder realisations during $5\cdot10^4$ MC sweeps. This
provides us with the working estimate for the critical temperature
for $L=12$, $T_{sim}(12)$. The process is continued until all the
lattice sizes are processed. The temperatures $T_{sim}(L)$
obtained in this way have been used for the final simulations of
$10^3$ disorder realisations from which all the main results are
driven. It is to note here that one has different possibilities to
define the critical temperature of a finite-size system. Along
with the definition used in this study, one can define $T_c(L)$ as
the temperature where the configurational average of $\chi$ has
its maximum. However the same value of $T_c$ is expected in the
thermodynamic limit and the same scaling behaviour holds when
approaching this limit.

The large clusters of correlated spins exist in the vicinity of
the critical point, that in turn leads to the effect of
critical slowing down \cite{HohHalp,Sokal}. As the result, the
relaxation time of the system increases dramatically. One can
estimate the typical relaxation time during the simulation by
observing the (configurationally dependent) autocorrelation
function \cite{AllenTild} for the potential energy (or,
alternatively, for some other characteristic of the system)
\begin{equation}
C({\tt t}) = \frac{\langle E({\tt t}_0)E({\tt t}_0+{\tt t})\rangle
- \langle E({\tt t}_0)\rangle \langle E({\tt t}_0)\rangle }
{\langle E({\tt t}_0)E({\tt t}_0)\rangle -\langle E({\tt
t}_0)\rangle \langle E({\tt t}_0)\rangle },
\end{equation}
where $E({\tt t})$ is the instant value for the energy at some
effective time ${\tt t}$ (in MC simulations ${\tt t}$ is measured
in MC sweeps) and ${\tt t}_0$ is some time origin. The $C({\tt
t})$ is averaged over different time origins in a course of
simulations. At times large enough $C({\tt t})$ decays
exponentially according to Debye law
\begin{equation}\label{tauE}
C({\tt t}) = \exp(-{\tt t}/\tau),
\end{equation}
more details can be found elsewhere \cite{Janke}. In (\ref{tauE})
$\tau$ has a meaning of the characteristic autocorrelation
time for a given disorder realisation. Besides the influence on
critical dynamics (which is beyond the scope of this study), the
critical slowing down has also some practical implications.
Firstly, the estimate of $\tau$ is vital for determination of the
required length of the simulation. Only the configurations
separated by a number of MC sweeps of order $\tau$ can be
considered as statistically independent, and the required length
of the simulation run should be measured in $\tau$ scale but not
in MC sweeps \cite{Janke}. Secondly, the critical slowing
down leads in Metropolis (or other local dynamics) algorithms to a
dramatic increase of $\tau$. This is, of course, due to a high
energy penalty required to flip a single spin (or pair of spins)
in a cluster of uniformly oriented neighbours. To overcome this
problem, a
number of cluster algorithms with non-local dynamics have been
suggested, with the Swendsen-Wang \cite{SW} and Wolff \cite{Wolff}
ones being most commonly used.

The second set of our simulations is targeting upon the detailed
comparison of autocorrelation times and of the efficiency of
different MC algorithms at a number of lattice sizes $L$ for the
model under consideration. To this end we performed the
simulations of 10-15 disorder realizations for each $L$ at the
temperatures $T_{sim}(L)$. We used three different algorithms, the
Metropolis, Swendsen-Wang and Wolff ones. The simulation length
was fixed to $10^5$ for all cases. For each given disorder
realisation the value of $\tau$ has been evaluated according to
(\ref{tauE}). These values were averaged then over all disorder
realisations. The results for the average autocorrelation time
$\tau_E$ are presented in tab.~1. The relaxation times are given
in MC sweeps, where one sweep corresponds to refreshing of each
spin. This means, that for the Wolff algorithm we followed the
standard procedure of the autocorrelation time normalisation
taking into account the average cluster size \cite{Wolff}. One can
see the different rate of pseudodynamics for all three algorithms
used and the autocorrelation time $\tau_E$ can be seen as a
measure of the minimal number of MC sweeps separating two adjacent
uncorrelated configurations. However, all three algorithms differ
significantly on a time spent for one MC sweep and we will be more
interested in the time required to generate the next uncorrelated
configuration. This will reflect the true speed of each algorithm.
The latter can be estimated from an inverse time ${\tt t}_{trial}$
spent on $N_{trial}$ MC sweeps and taking into account that only
$N_{trial}/\tau_E$ of these are uncorrelated
\begin{equation}
\lambda = \frac{N_{trial}}{{\tt t}_{trial}\tau_E}.
\end{equation}

\begin{table}[htb]
\begin{center}
\begin{tabular}{|l|l|l|l|l|l|l|l|l|}
\hline
L&
10   & 12   & 16   & 24   & 32   & 48   & 64   & 96 \\
\hline $\tau_{E}$, Metropolis&
5.79 & 7.87 & 12.9 & 26.5 & 44.9 & 93.1 & 160 & 336 \\
\hline $\tau_{E}$, Swendsen-Wang&
3.55 & 3.98 & 4.69 & 5.54 & 6.10 & 7.50 & 8.55 & 10.7 \\
\hline $\tau_{E}$, Wolff&
1.18 & 1.24 & 1.31 & 1.46 & 1.54 & 1.71 & 1.81 & 1.98 \\
\hline
\end{tabular}
\end{center}
\caption{The energy autocorrelation time $\tau_E$ for different
lattice sizes $L$ measured in MC sweeps (see the text for details).}
\end{table}

\begin{figure}[h]
\leavevmode
\begin{center}
\epsfxsize=8cm \epsffile{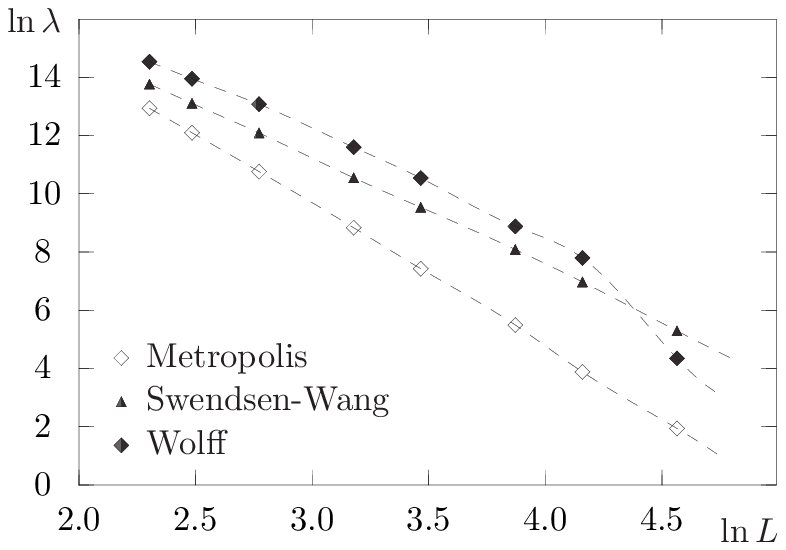}
\end{center}
\caption{\label{fig1} Log-log plot for relative speed measured as on inverse
time spent on generation of one uncorrelated configuration for
lattice sizes $L=10-96$ and three different MC
algorithms}
\end{figure}

To this end we performed short trial simulations for 10 disorder
realisations each of which was well equilibrated, with
$N_{trial}=10^3$. The simulations were performed for all lattice
sizes $L=10-96$ and using all three MC algorithms. The
results are presented in fig.~\ref{fig1}.
One can conclude the following. Both cluster algorithms have
significantly higher real speed than the Metropolis one for all
lattice sizes. Concerning the larger lattices examined both
cluster algorithms demonstrate comparable speed with the Wolff one
being slightly faster for $L<80$. At $L\sim 80$ the speed of both
cluster algorithms equals, and for the larger lattice sizes the
Swendsen-Wang algorithm looks more preferable (however, we have
not done any simulations beyond $L=96$ to check this). According
to these findings we can conclude that for the model and lattice
sizes examined in this study the Wolff algorithm is the most
preferable one and therefore it was used to obtain all the main
results in this study. We also should note that our realisation of
the Metropolis algorithm does not use any special tricks to
speed-up the simulation of the Ising model (energy maps,
multi-spin coding, etc.). Therefore, our estimate of a true speed
of Metropolis algorithm is subject to the particular code being
used.

The third set of our simulations aimed on the evaluation of
the critical indices and the extrapolation of the critical
temperature $T_c(\infty)$ for the infinite system.
Along with the existing MC estimates of the
critical indices of diluted 3D Ising model
(see e.g. Refs. \cite{Pelissetto02,Folk03} for reviews)
for the sake of consistency  we performed our
independent evaluation of these quantities. To this
end the MC simulations of $10^3$ disorder realisations
were performed for lattice sizes $L=10,12,16,24,32,48,64,96$
using Wolff algorithm. The simulation length was chosen
equal to $250\tau_E$ for the equlibration and $10^3\tau_E$
for the production runs. These numbers were chosen to balance
the errors originated from the limited number of disorder
realisations $N_{dis}$ and the finite length $N_{MC}$
of the simulations, both terms appearing in the total
error estimate for some quantity $A$ \cite{Janke}
\begin{equation}\label{deltaA}
\Delta A = \left( \frac{\sigma_{\langle A\rangle}^2}{N_{dis}} +
(1+2\tau_A)\frac{\sigma_A^2}{N_{dis}N_{MC}} \right)^{1/2},
\end{equation}
where
\begin{equation}
\sigma_{\langle A\rangle} = \left(
  \overline{\langle A^2\rangle} - \overline{\langle A\rangle}^{\,2}
  \right)^{1/2}
\end{equation}
is the standard deviation of the set of $N_{dis}$ averages
$\langle A\rangle$ calculated within each disorder realisations,
and
\begin{equation}
\sigma_{A} = \left(\overline{\langle A^2\rangle - \langle
A\rangle^2 }\right)^{1/2}
\end{equation}
is the standard deviation of the whole set of $N_{dis}N_{MC}$ data
for the quantity $A$. The $\tau_A$ in Eq.(\ref{deltaA}) is the
integrated aucorrelation time $\tau^{int}_A$ \cite{Janke} which,
in general differs from the exponentional one $\tau^{exp}_A$
introduced in (\ref{tauE}) in the case of the energy. However, for
most of realistic models $\tau^{int}_A \leq \tau^{exp}_A$
\cite{Janke} so that the use of $\tau^{exp}_A$ only which we
employ in this study is reasonable.

\begin{figure}[h]
\leavevmode
\begin{center}
\epsfxsize=8cm \epsffile{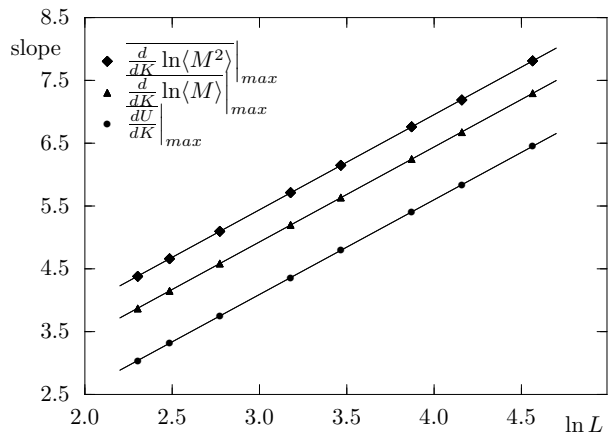}
\end{center}
\caption{\label{fig2} Log-log plots for the maximum values of the
configurationally averaged derivatives of Binder cumulant
\protect(\ref{maxval}) (discs), $\ln \langle M \rangle$
\protect(\ref{logm}) (triangles) and $\ln \langle M^2 \rangle$
\protect(\ref{logmsq}) (diamonds)} \end{figure}

In evaluation of the critical indices
we followed the standard finite-size scaling scheme
\cite{FerLand}. According to it, the Binder's cumulant
\begin{equation}\label{UBind}
U = 1 - \frac{\langle M^4\rangle }{3\langle M^2\rangle ^2}
\end{equation}
can be evaluated as a function of a coupling $K$ for each disorder
configuration and the maximum value for the slope of this function
varies with the system size as $L^{1/\nu}$ anywhere in the
critical region. In the terms of temperature derivative one has
\begin{equation}\label{maxval}
\left.\frac{dU}{dK}\right|_{max} =
-T^2\left.\frac{dU}{dT}\right|_{max} \sim aL^{1/\nu},
\end{equation}
where the histogram reweighting technique was used to evaluate the
cumulant $U$ in the neighbourhood of $T_c(L)$ and the numerical
derivation was employed. Here and thereafter the observables (e.
g. (\ref{maxval})-(\ref{logmsq})) are averaged over disorder
realisation. The inverse values for $\nu$ can be found by the
linear interpolation of the log-log plot for the values
$\left.\overline{\frac{dU}{dK}}\right|_{max}$ vs $L$ (see,
fig.~\ref{fig2}). The same finite size scaling behaviour is valid for the
number of logarithmic derivatives for the powers of the
magnetisation \cite{FerLand}. These can be used as additional
estimates for the critical index $\nu$, and we evaluated two of
them, namely
\begin{eqnarray}\label{mmsq}
\overline{\frac{d}{dK}\ln\langle M\rangle } &=&
 \overline{-T^2 \frac{d}{dT}\ln\langle M\rangle} ,\label{logm}\\
\overline{\frac{d}{dK}\ln\langle M^2\rangle } &=&
 \overline{-T^2 \frac{d}{dT}\ln\langle M^2\rangle} .\label{logmsq}
\end{eqnarray}
The log-log plots for these derivatives are shown in fig.~\ref{fig2}, too.

First of all we would like to stress that all the data points
presented in this figure interpolate extremely well by the
appropriate linear dependencies. This can be seen as an other
proof that the number of disorder realisations and the simulation
length that have been used are reasonable. The linear
interpolation for the curves presented in fig.~\ref{fig2} leads to the
values presented in tab.~2.

\begin{table}
\begin{center}
\begin{tabular}{|l|l|l|}
\hline
interpolation of & $1/\nu$ & $\nu$ \\
\hline $\left.\overline{\frac{dU}{dK}}\right|_{max}$
      & $1.507 \pm 0.008$ & $0.664 \pm 0.004$\\
$\left.\overline{\frac{d}{dK}\ln\langle M\rangle }\right|_{max}$
      & $1.512 \pm 0.001$ & $0.661 \pm 0.001$\\
$\left.\overline{\frac{d}{dK}\ln\langle M^2\rangle}\right|_{max}$
  & $1.514 \pm 0.001$ & $0.660 \pm 0.001$\\
\hline
\end{tabular}
\end{center}
\caption{The values for the critical index $\nu$ obtained via linear
interpolation of the log-log data for the slope of Binder's cumulant
and logarithmic derivatives of $\langle M \rangle$ and
$\langle M^2 \rangle$}
\end{table}

The average of all the three results for the index $\nu$ shown in
tab.~2 yields
\begin{equation}\label{resultnu}
\nu = 0.662 \pm 0.002
\end{equation}
Obtained value for $\nu$ distingtly differs from the theoretical
estimate of the corresponding exponent of the pure 3D Ising model:
$\nu = 0.6304(13)$ \cite{Guida98}. However it is less
than the asymptotic value of $\nu$ for the random 3D Ising model
as obtained by the field-theoretical RG approach $\nu = 0.678(10)$
\cite{Pelissetto00} and by the MC simulations $\nu = 0.6837(53)$
\cite{Ballesteros98}, $\nu =0.683(3)$ \cite{Calabrese03}. This is
an evidence of the  fact that for the spin concentrations and
lattice sizes chosen here  the system still crosses over to the
asymptotic regime. Note, that close estimates for $\nu$ were
found at similar system parameters (and neglecting
correction-to-scaling terms) in the MC simulation of the random
site 3D Ising model: $p=0.9, L=64\div 128, \nu = 0.6644(15)$
\cite{Ballesteros98}; as well as of the random bond 3D Ising model:
$p_{bonds}=0.7, \nu= 0.660(10)$ \cite{Berche04}.

\begin{figure}[h]
\leavevmode
\begin{center}
\epsfxsize=8cm \epsffile{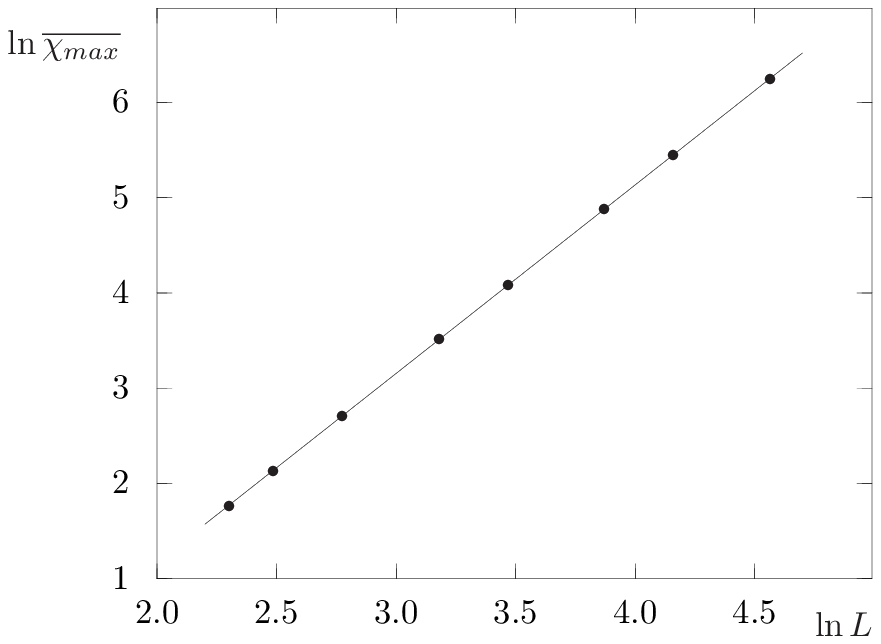}
\end{center}
\caption{\label{fig3} Log-log plot for the maximum value of the
susceptibility $\overline{\chi_{max}} $ vs the linear size of the
system $L$ in the critical region}
\end{figure}

\begin{figure}[h]
\leavevmode
\begin{center}
\epsfxsize=8cm \epsffile{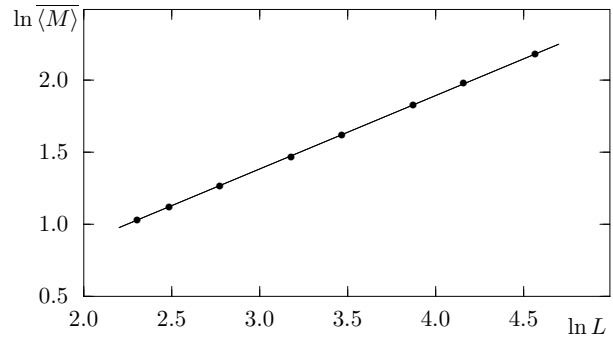}
\end{center}
\caption{\label{fig4} Log-log plot for the magnetization $\overline{\langle
M\rangle}$ vs the linear size of the system $L$ in the critical
region}
\end{figure}

The standard finite-size scaling concepts have
been also employed for the evaluation of the other critical
indices, particularly these for the susceptibility
$\gamma$ and for the magnetisation $\beta$. Due to this
scheme \cite{FerLand} for finite size system of dimension
$L$ in the critical region one has
\begin{equation}\label{chiM}
\overline{\chi_{max}} \sim \Gamma L^{\gamma/\nu},\;\;
\overline{\langle M\rangle} \sim B L^{-\beta/\nu},
\end{equation}
where $\Gamma$ and $B$ are the critical amplitudes and the
correction-to-scaling terms have been omitted. In (\ref{chiM})
$\overline{\chi_{max}}$ is the averaged maximum value for the
susceptibility that is achieved at some temperature and
$\overline{\langle M\rangle} $ is the averaged value of the
magnetisation at the same temperature. We employed the following
scheme. For each disorder realisation the temperature $T^*$ was
found where the $\chi$ achieves its maximal value $\chi_{max}$.
Then the magnetisation $\langle M\rangle$ is evaluated at the same
$T^*$. Afterwards, the values of $\chi_{max}$ and $\langle
M\rangle$ have been averaged over all disorder realisations to be
used for fitting the relations (\ref{chiM}). The results of
fittings are presented in figs.~\ref{fig3},\ref{fig4} in a form of
the log-log plots for the $\overline{\chi_{max}} $ and for
$\overline{\langle M\rangle}$ vs $L$, respectively.

One should note near perfect linear fits achieved in both cases,
leading to the following results for $\gamma$ and $\beta$
\begin{eqnarray} \label{gam}
\gamma/\nu &=& 1.986 \pm 0.001, \\
\beta/\nu  &=& 0.509 \pm 0.001, \\ \label{gamma}
\gamma     &=& 1.314 \pm 0.004, \\
\label{beta} \beta      &=& 0.337 \pm 0.001.
\end{eqnarray}
For the reasons explained above, taking exponent $\nu$ as an
example, the values (\ref{gam})-(\ref{beta}) slightly differ from
their asymptotic counterparts $\beta=0.3546(28),\gamma =1.342(10)$
\cite{Ballesteros98}. They are in a good agreement with the
exponents of the 3D Ising model with random bond dilution:
$\beta/\nu=0.515(5), \gamma/\nu=1.97(2)$ \cite{Berche04} giving
one more proof that both models belong to the same universality
class.

\begin{figure}[h]
\leavevmode
\begin{center}
\epsfxsize=8cm \epsffile{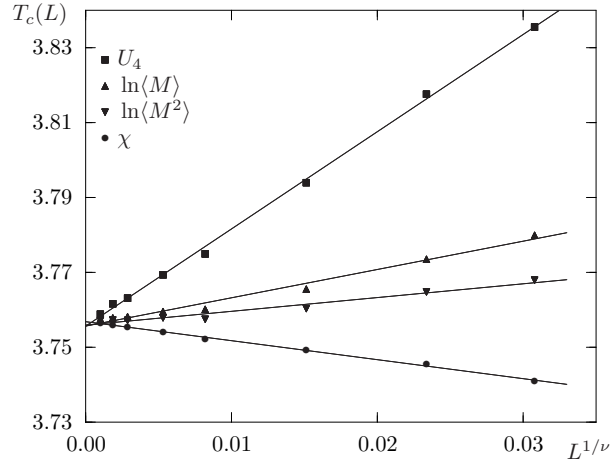}
\end{center}
\caption{\label{fig5} The linear fits for the critical temperatures
$T^{i}_c(L)$ vs the scaled system size $L^{-1/\nu}$ for the
lattice sizes from $L=10$ to $L=96$, the index $i$ corresponds to
the maximum values for different quantities, see notations in the
figure}
\end{figure}

With the value for the critical index $\nu$ being evaluated in
(\ref{resultnu}) one can obtain the estimate for the critical
temperature for the infinite system $T_c(\infty)$. It is related
to the critical temperatures of the finite system of linear
dimension $L$, $T_c(L)$, via the relation (\ref{Tc}). The values for
$T_c(L)$ were estimated during the same set of simulations that
have been used for the evaluation of critical indices.
For each disorder realisation we found the sets
of temperatures $T^{i}_c(L)$ where the maximum values for the
susceptibility (for the $i=1$ case) and for the derivatives
(\ref{maxval})-(\ref{logmsq}) (the cases $i=2,3,4$, respectively)
are achieved. The data sets $T^{i}_c(L)$ should be plotted then vs
the scaled system size $L^{-1/\nu}$ and, ideally, for all $i$
should extrapolate at $L\rightarrow\infty$ to the same value of
$T_c(\infty)$. The results are shown in fig.~\ref{fig5} where we used the
data for all the lattice sizes from $L=10$ to $L=96$. As the result
of the extrapolation procedure we obtain
\begin{equation}\label{Tinfty}
T^{i}_c(\infty) = \left\{
\begin{array}{ll}
3.756931\pm 0.000058, & \chi|_{max} \\
3.756783\pm 0.000243, & dU/dK|_{max}\\
3.756481\pm 0.000086, & d\ln\langle{M}\rangle/dK|_{max} \\
3.756543\pm 0.000083, & d\ln\langle{M^2}\rangle/dK|_{max}
\end{array}
\right.
\end{equation}

One should note very good convergence of the
extrapolated values for $T^{i}_c(\infty)$ found by examining of
different quantities and given in Eq.(\ref{Tinfty}). The final
estimate for the critical temperature of the infinite system
$T_c(\infty)$ can be found as an average on all four numbers
in Eq.(\ref{Tinfty})
\begin{equation}\label{Tinftyres}
T_c(\infty) = 3.7566845\pm 0.0001175,\,\,\,\,\,
\beta_c=T_c^{-1}(\infty)=0.2661922\pm 0.0000083.
\end{equation}

As it is clear already from simple mean-field arguments (see Eq.
(3)) our estimate for $T_c$ made for the spin concentration
$p=0.85 $ should lay in between corresponding data for $p=0.8$ and
$p=0.9$. Indeed, the most recent estimates read:
$\beta_c(p=0.8)=0.2857368(52)$; $\beta_c(p=0.9)=0.2492880(35)$
\cite{Ballesteros98}, $\beta_c(p=0.8)=0.2857447(24)$
\cite{Calabrese03}.

\begin{figure}[h]
\leavevmode
\begin{center}
\epsfxsize=8cm \epsffile{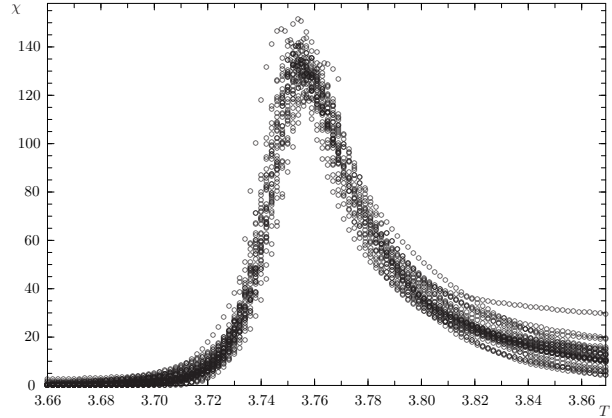}
\end{center}
\caption{\label{fig6}
The example of the distribution for the susceptibility
vs temperature curves for different disorder realisations for
the system size $L=48$}
\end{figure}

In this study we have calculated the universal ratio
$\Gamma^+/\Gamma^-$ for the magnetic susceptibility in the
critical region
\begin{equation}\label{chipm}
\overline{\chi} = \left\{
\begin{array}{ll}
\Gamma^{+}t^{-\gamma}, & T>T_c \\
\Gamma^{-}t^{-\gamma}, & T<T_c.
\end{array}
\right.
\end{equation}
The singularity of the susceptibility (\ref{chipm}) is observed
for the infinite system only, at finite system size $L$ it is
rounded-off with the finite maximum value. At each given
system size $L$ there will be only a finite temperature interval
where the susceptibility curve overlaps with that for the
infinite system. As $T$ approaches the $T_c(\infty)$ the
finite-size curve deviates from the infinite one. As the
result, each estimate for the critical amplitudes made at
different system sizes will be valid in certain temperature
range. The other complication for the disordered systems is
the presence of some distribution of $\chi$ curves obtained
for different disorder realisations (the example for the system
size $L=48$ is presented in fig.~\ref{fig6}).
We employed the following method for the evaluation of critical
amplitudes $\Gamma^+$ and $\Gamma^-$. The critical temperature for
the infinite system $T_c(\infty)$ was taken as the central one and
then the grid of temperatures around it was considered. The
gridness was chosen to be even in scaled temperature units
$L^{1/\nu}|t|$, where
$t=K_c(\infty)-K=\frac{(T-T_c(\infty))}{TT_c(\infty)}$. At each
grid temperature the separate simulation of $10^2$ disorder
realisations was performed of the length $10^3\tau_{E}$ and the
value for the susceptibility was evaluated. One could, in
principle, use the histogram reweighting technique to evaluate the
intermediate temperatures as well, but this will require the study
of histograms validity. In this calculation we opted for
more straightforward approach using the separate simulation
runs and haven't employed histogram reweighting.

\begin{figure}[h]
\leavevmode
\begin{center}
\epsfxsize=8cm \epsffile{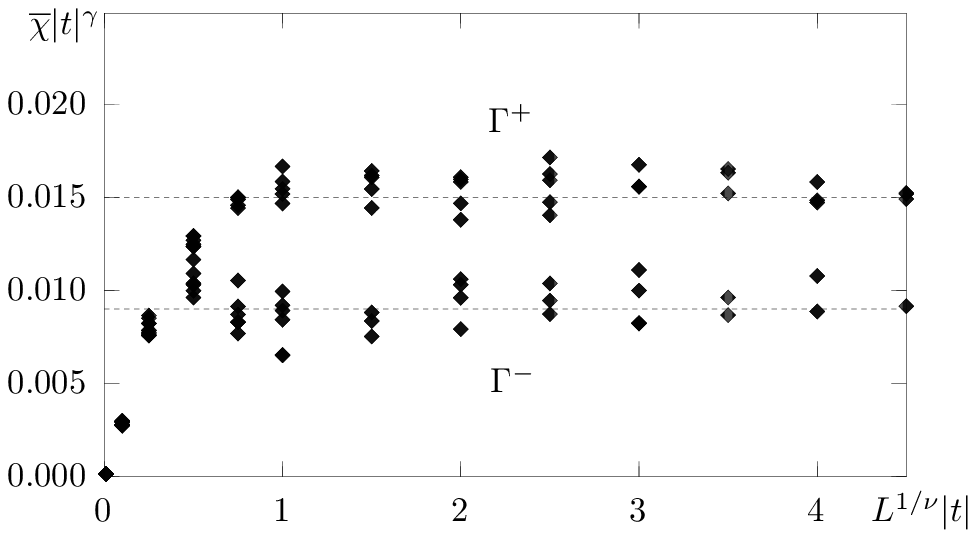}
\end{center}
\caption{\label{fig7} The values for the critical amplitudes for the
susceptibility $\Gamma^+$ and $\Gamma^-$ at $T>T_c(\infty)$ and
$T<T_c(\infty)$, respectively}
\end{figure}

The values for the critical amplitudes $\Gamma^+$ and $\Gamma^-$
can be obtained by plotting the scaled susceptibility
$\overline{\chi}|t|^\gamma$. The data are presented in
fig.~\ref{fig7} and for the evaluation of the ratio
$\Gamma^+/\Gamma^-$ the data points with $L^{1/\nu}|t|<1$ have
been ignored, since they are in the FSS regime .
This gives the following value for the amplitudes ratio
\begin{equation}\label{amp}
\Gamma^+/\Gamma^- = 1.67 \pm 0.15
\end{equation}
It compares very well with the recent result
$\Gamma^+/\Gamma^-=1.62 \pm 0.1$ \cite{Berche04}
obtained by MC simulations of bond-diluted Ising model
at bond dilution $p_{bond}=0.7$. The
theoretical prediction obtained by the field-theoretical
renormalization group calculations in three-loop approximation
reads: $\Gamma^+/\Gamma^- = 3.05(32)$ \cite{BerShpot}. As a reason
of the discrepancy between estimates (\ref{amp}), \cite{Berche04}
and \cite{BerShpot} one can
mention the shortness and  bad convergence of the series for
$\Gamma^+/\Gamma^-$, as recognized already in
\cite{BerShpot}. However further analysis is required to get
a definite answer. In any case, both
theoretical and MC estimates of the critical amplitudes ratio
certainly differ from those of the pure 3D Ising model:
$\Gamma^+/\Gamma^- = 4.70 \div 4.95$
(see \cite{Caselle97} and references therein).

\section{Conclusions and outlook}\label{III}

In this study we performed computer simulations of the 3D Ising
model with random site dilution in critical region. Despite the
number of simulational studies of this model being already
performed, some aspects of its critical behaviour and
simulational details are still awaiting for more
detailed analysis. One of these to mention is the applicability
and the effectiveness of different Monte Carlo algorithms, the
topic being studied previously for the pure models mainly.
However in the case of disordered models the impurities may
have a profound effect on system clusterisation which, in turn,
is exploited in cluster methods.

We examined the most commonly used Metropolis, Swendsen-Wang and
Wolff algorithms and evaluated both the energy autocorrelation
times and the respective speed of algorithms in generating
statistically independent configurations. We found that for the
linear lattice sizes up to $L=100$ the Wolff algorithm is more
preferable to use and hence it was applied in our study.

We used the workstations cluster of the ICMP currently equipped by
the Athlon MP--2200+ processors. The typical simulation times
range from 0.1 seconds per 1000 MC sweeps for the smallest lattice
size $L=10$ (Wolff algorithm) to about 953 seconds for the largest
lattice size $L=96$ (Metropolis algorithm) per one disorder
realisation.

Following the standard ideas of the finite size scaling,
the critical indices have been calculated. In this way
we complemented the existing results. Based on the fact
that the correction-to-scaling terms appear to be
particularly small \cite{Ballesteros98} in the region of
spin concentration $p\sim 0.8$, we have performed the simulation
for the concentration $p=0.85$ only.
Obtained values for the critical exponents (\ref{resultnu}),
(\ref{gamma}),(\ref{beta}) slightly differ from their asymptotic
values as obtained by
the field-theoretical RG approach \cite{Pelissetto00} and by
recent MC simulations \cite{Ballesteros98}, \cite{Calabrese03}.
This is the consequence of the  fact that for the spin concentrations and
lattice sizes chosen here  the system still crosses over to the
asymptotic regime.

Contrary to the critical indices, the values of critical
amplitude ratios received so far less attention. In our study, we
get an estimate for the magnetic susceptibility critical
amplitude ratio $\Gamma^+/\Gamma^- = 1.67 \pm 0.15$. This value
is in a good agreement with the recent MC analysis of the
random-bond Ising model \cite{Berche04} giving further support
that both random-site and random-bond dilutions lead to the same
universality class.

The comparative analysis of different MC algorithms
(including Metropolis, Swendsen-Wang and Wolff ones) which
is a subject of this study will be used in our future work on
influence of different types of structural disorder on criticality.

\section*{Acknowledgements}

We acknowledge useful discussions with
Wolfhard Janke and Christophe Chatela\-in.
This work was supported by the French-Ukrainian cooperation
"Dnipro" program\-me. Work of Yu.H. was supported
in part by the Austrian Fonds zur F\"{o}rderung der
wissenschaftlichen Forschung, project No. 16574-PHY.

It is our special pleasure and pride to acknowledge
wonderful hospitality and high spirit of people in Kyiv,
where several of the authors met in December days of 2004 and
where a part of this paper was finalized.


\begin{thebibliography}{17}




\bibitem{Harris}
Harris A.B. J. Phys. C, 1974, vol.~7, p.1671.

\bibitem{Pelissetto02}
Pelissetto A., Vicari E. Phys. Rep, 2002, vol.~368, p.~549.

\bibitem{Folk03}
Folk R., Holovatch Y., Yavors'kii T. Uspekhi Fiz. Nauk, 2003,
vol.~173, p.~175., [Phys. Usp. 2003, vol.~46, p.~175].

\bibitem{Ballesteros98}
Ballesteros H.G., Fern\'{a}ndez L.A., Marti\'{n}-Mayor V.,
Mu\~{n}oz Sudure A. Phys. Rev.~B, 1998, vol.~58, p.~2740.

\bibitem{Barber}
Barber M.N. Finite-size scaling. -- In: Phase Transitions and
Critical Phenomena, eds. C.Domb and J.Lebowitz, Academic, New
York, 1983, vol.8.

\bibitem{FerLand}
Ferrenberg A.M., Landau D.P. Phys. Rev.~B, 1991, vol.~44, No.~10,
p.5081-5091.

\bibitem{Metr}
Metropolis N. {\it et al}  J. Chem. Phys., 1953, vol.~21, No.~6,
p.~1087-1092.

\bibitem{FerSw}
Ferrenberg A.M., Swendsen R.H. Phys. Rev. Lett., 1988, vol.~61,
No.~23, p.2635-2638.

\bibitem{HohHalp}
Hohenberg P.C., Halperin B.I. Rev. Mod. Phys., 1977, vol.~49,
No.~3, p.~435-479.

\bibitem{Sokal}
Sokal A.D. MC Methods in Statistical Mechanics:
Foundations and New Algorithms, Cours de Troisi\'{e}me Cycle de la
Physique en Suisse Romande, Lausanne, June 1989.

\bibitem{AllenTild}
Allen M.P., Tildesley D.J. Computer Simulation of Liquids.
Clarendon Press, Oxford, 1987.

\bibitem{Janke}
Janke W. Statistical analysis of simulations: data correlations
and error estimation. In: Quantum Simulations of Complex Many-Body
Systems: From Theory to Algorithms, Lecture Notes, J.Grotendorst,
D.Marx, A.Muramatsu, eds., John von Neumann Institute for
Computing, J\"{u}lich, NIC Series, 2002, vol.~10, p.~423-445.

\bibitem{SW}
Swendsen R.H., Wang J.-S. Phys. Rev. Lett., 1987, vol.~58, No.~2,
p.~86-88.

\bibitem{Wolff}
Wolff U. Phys. Rev. Lett., 1989, vol.~62, No.~4 , p.~361-364.

\bibitem{SW2002}
Wang J.-S., Kozan O., Swendsen R.H. Phys. Rev. E {\bf 66}, 057101
(2002)

\bibitem{Berche04}
Berche P.E., Chatelain C., Berche B., Janke W. Eur. Phys. J. B,
vol.~38, p.~463 (2004).

\bibitem{BerShpot}
Bervillier C., Shpot M. Phys. Rev.~B, 1992, vol.~46, No.~2,
p.~955-968.


\bibitem{Calabrese03}
Calabrese P., Marti\'{n}-Mayor V., Pelissetto A., Vicari E. Phys.
Rev.~B, 2003, vol.~68, p.~036136.

\bibitem{Guida98}
Guida R., Zinn-Justin J. J. Phys. A, 1998, vol.~31, p.~8103.

\bibitem{Pelissetto00}
Pelissetto A., Vicari E. Phys. Rev.~B, 2000, vol.~62, p.~6393.

\bibitem{Caselle97}
Caselle M., Hasenbusch M. J. Phys. A, 1997, vol.~30, p.~4963.

\end{thebibliography}
\end{document}